\documentclass[sigconf]{acmart}

\usepackage{tabularx}
\usepackage{url}
\usepackage{booktabs}
\usepackage{tabularx}
\usepackage{float}
\usepackage{array}
\usepackage{tcolorbox}
\usepackage{listings}
\usepackage{tikz}
\usepackage{seqsplit}

\makeatletter
\newcommand{\textbfit}[1]{\textbf{\textit{#1}}}
\makeatother

\newcommand{\emptycirc}{%
    \tikz[baseline={(0,-0.7ex)}]{%
        \node[shape=circle,draw,inner sep=0pt,minimum size=1.5ex] (char) {};
    }%
}
\newcommand{\halfcirc}{%
    \tikz[baseline={(0,-0.7ex)}]{%
        \node[shape=circle,draw,inner sep=0pt,minimum size=1.5ex] (char) {};
        \fill[black] (char.center) -- (char.north) arc[start angle=90,end angle=270,radius=0.75ex] -- cycle;
    }%
}
\newcommand{\fullcirc}{%
    \tikz[baseline={(0,-0.7ex)}]{%
        \node[shape=circle,draw,fill=black,inner sep=0pt,minimum size=1.5ex] (char) {};
    }%
}
\newcommand{\tikzxmark}{%
\tikz[scale=0.23] {
    \draw[line width=0.7,line cap=round] (0,0) to [bend left=6] (1,1);
    \draw[line width=0.7,line cap=round] (0.2,0.95) to [bend right=3] (0.8,0.05);
}}
\newcommand{\tikzcmark}{%
\tikz[scale=0.23] {
    \draw[line width=0.7,line cap=round] (0.25,0) to [bend left=10] (1,1);
    \draw[line width=0.8,line cap=round] (0,0.35) to [bend right=1] (0.23,0);
}}

\begin{document}

\title[Breaking Bluetooth Security Abusing Silent Automatic Pairing]{Stealtooth: Breaking Bluetooth Security Abusing Silent Automatic Pairing}

\author{Keiichiro Kimura}
\authornotemark[1]
\affiliation{%
  \institution{Kobe University}
  \city{Kobe}
  \country{Japan}
}
\email{233t225t@gsuite.kobe-u.ac.jp}

\author{Hiroki Kuzuno}
\affiliation{%
  \institution{Kobe University}
  \city{Kobe}
  \country{Japan}
}
\email{kuzuno@port.kobe-u.ac.jp}

\author{Yoshiaki Shiraishi}
\affiliation{%
  \institution{Kobe University}
  \city{Kobe}
  \country{Japan}
}
\email{zenmei@port.kobe-u.ac.jp}

\author{Masakatu Morii}
\affiliation{%
  \institution{Kobe University}
  \city{Kobe}
  \country{Japan}
}
\email{mmorii@kobe-u.ac.jp}

\renewcommand{\shortauthors}{Kimura et al.}
\begin{abstract}

Bluetooth is a pervasive wireless communication technology used by billions of devices for short-range connectivity. The security of Bluetooth relies on the pairing process, where devices establish shared long-term keys for secure communications. However, many commercial Bluetooth devices implement automatic pairing functions to improve user convenience, creating a previously unexplored attack surface.

We present \textbf{Stealtooth}, a novel attack that abuses unknown vulnerabilities in the automatic pairing functions in commercial Bluetooth devices to achieve completely silent device link key overwriting. The Stealtooth attack leverages the fact that Bluetooth audio devices automatically transition to pairing mode under specific conditions, enabling attackers to hijack pairing processes without user awareness or specialized tools. We also extend the attack into the \textbf{MitM Stealtooth} attack, combining automatic pairing abuse with power-saving mode techniques to enable man-in-the-middle attacks.

We evaluate the attacks against 10 commercial Bluetooth devices from major manufacturers, demonstrating widespread vulnerabilities across diverse device types and manufacturers. Our practical implementation requires only commodity hardware and open-source software, highlighting the low barrier to entry for attackers.

We propose defenses both device and protocol levels, including enhanced user notifications and standardized automatic pairing guidelines. Our findings reveal a critical tension between security and usability, showing that current automatic pairing implementations create systematic vulnerabilities. We responsibly disclosed our findings to affected vendors, with several already releasing patches.
\end{abstract}

\keywords{Do, Not, Us, This, Code, Put, the, Correct, Terms, for,
  Your, Paper}

\maketitle

\section{Introduction}

Bluetooth has become one of the most ubiquitous wireless communication technologies, with billions of devices relying on it for short-range connectivity across personal computing, IoT, peripheral, and wearable applications\cite{bluetoothsig2023market,bluetoothsig2024market,bluetoothsig2025market}. The widespread adoption of Bluetooth across diverse device types—from smartphones and laptops to headsets, keyboards, and industrial sensors—has made it a critical component of modern digital infrastructure\cite{icondsc2019ble,spw2022vehicles,sensor2012ble,sensor2020ble,sensor2025ble}. As these devices increasingly handle sensitive data including audio communications, personal files, and control commands, the security of Bluetooth connections has become paramount.

The security architecture of Bluetooth relies fundamentally on the pairing process, where devices establish a shared long-term key (link key) that serves as the foundation for all subsequent secure communications\cite{ccs2023bluffs,woot2020blesa}. The pairing process, typically requiring user confirmation, creates a trusted relationship and secure sessions between devices that should prevent unauthorized access and impersonation attacks. Once paired, devices can automatically reconnect and establish secure sessions without repeated user intervention, enabling the seamless connectivity that users expect from Bluetooth technology.

However, the convenience comes with inherent security risks. The automatic reconnection functionality, while improving user experience, creates potential attack vectors that have not been thoroughly investigated by the security community. Many commercial Bluetooth devices, particularly audio peripherals like headsets and speakers, implement automatic pairing functions designed to simplify the user experience by reducing the need for manual intervention during connection establishment. The automatic pairing functions allow devices to transition automatically into pairing states under certain conditions, such as when attempting to reconnect to previously paired devices.

Prior work has extensively investigated various Bluetooth security vulnerabilities, revealing critical weaknesses in different aspects of the protocol. Notable attacks such as BIAS (Bluetooth Impersonation AttackS) \cite{sp2020bias} and Blacktooth \cite{ccs2022blacktooth} have primarily focused on abusing authentication weaknesses, role switching vulnerabilities, and encryption key negotiation flaws in the core Bluetooth specification. The BIAS attack abuses unidirectional authentication procedures to enable device impersonation, while Blacktooth leverages the flexibility of Master/Slave role assignments to establish unauthorized connections. Additionally, attacks like KNOB \cite{usenix2019knob} have demonstrated vulnerabilities in encryption key negotiation processes that can significantly weaken communication security.

Beyond Bluetooth-specific attacks, the broader domain of wireless communication security has seen significant research attention in recent years. Studies on wireless device pairing vulnerabilities have identified various attack vectors, including acoustic eavesdropping during pairing processes \cite{ieeetifs2013pairing} and method confusion attacks that exploit differences in authentication methods \cite{ndss2025mc, sp2021mc}. Research on WiFi impersonation detection \cite{ieeetifs2018wifi} and radio frequency fingerprinting \cite{ieeetifs2022lora} has also contributed to our understanding of wireless communication security challenges.

However, despite this extensive body of work, prior attacks have not analyzed the security implications of automatic pairing functions that are increasingly common in consumer Bluetooth devices. This represents a significant gap in the current understanding of Bluetooth security, as automatic functionalities may introduce novel attack vectors that bypass traditional security assumptions about user involvement in the pairing process. Unlike specification-level vulnerabilities that affect all compliant devices, automatic pairing behaviors vary significantly across manufacturers and device types, creating a diverse and previously unexplored attack surface.

Our work addresses this gap by investigating automatic pairing vulnerabilities for the first time, revealing a critical but overlooked attack surface in modern Bluetooth implementations. In this paper, we present \textbf{Stealtooth}, a novel attack that abuses previously unexplored vulnerabilities in Bluetooth's automatic pairing functions. The Stealtooth attack achieves \textit{completely silent} and \textit{stealthy} link key overwriting by abusing the automatic pairing functions implemented in commercial Bluetooth devices. The Stealtooth attack leverages the fact that many Bluetooth audio devices automatically transition to pairing functionalities when they cannot establish connections with previously paired devices, creating an opportunity for attackers to hijack these pairing processes without any user awareness.

We demonstrate that the Stealtooth attack can be extended into a man-in-the-middle (MitM) attack, called the \textbf{MitM Stealtooth} attack, by combining it with existing techniques that abuse Bluetooth power-saving modes. The MitM Stealtooth attack enables attackers to position themselves between communicating devices and intercept, modify, or relay communications while maintaining the appearance of normal operation to the victims.

The implications of our findings are \textit{significant} for the broader Bluetooth ecosystem. The Stealtooth attacks work against commercially available devices without requiring any specialized hardware or deep protocol knowledge, making them practical threats that could be deployed by adversaries with minimal technical skills. Furthermore, the silent nature of the attacks makes them particularly dangerous, as users have no indication that their devices have been compromised.

We summarize our main contributions as follows:

\begin{itemize}
    \item We unveil and demonstrate a \textit{novel} vulnerability in Bluetooth automatic pairing functions that enables completely silent link key overwriting and propose the \textbf{Stealtooth} attack, which abuses the new vulnerabilities to establish malicious Bluetooth sessions with victim devices. The Stealtooth attack requires \textit{no} specialized hardware, protocol manipulation, or user interaction, making it highly practical and deployable by adversaries with minimal technical skills.
    \item We also present the \textbf{MitM Stealtooth} attack, which extends the Stealtooth attack into a MitM attack by combining automatic pairing mode abuse with existing power-saving mode techniques. The MitM Stealtooth attack enables attackers to intercept, modify, and relay communications between victims while maintaining operational transparency.
    \item We conduct a comprehensive evaluation of the Stealtooth attacks against 10 commercial Bluetooth devices from major manufacturers, demonstrating the widespread nature of automatic pairing vulnerabilities. Our evaluation covers diverse device types including headphones, earbuds, and speakers from Sony, Anker, Google, Xiaomi, and other manufacturers. We also propose practical defenses to mitigate the automatic pairing vulnerabilities and discuss the limitations of our current attack implementation.
\end{itemize}

\paragraph{\textbf{Ethical Consideration and Responsible Disclosure:}}
This work investigates unknown threats to widespread technologies and proposes defenses. All experiments were conducted in-house; no external devices were attacked. We responsibly disclosed our findings to the related manufacturers. Several manufacturers acknowledged our findings. Sony has already released patches for the vulnerable devices we reported, with an acknowledgment to us.

The remainder of this paper is organized as follows: Section 2 provides background on Bluetooth technology and connection establishment processes. Section 3 defines our threat model and attack assumptions. Section 4 presents the detailed design of both the Stealtooth and MitM Stealtooth attacks. Section 5 describes our implementation approach. Section 6 presents comprehensive evaluation results against real-world devices. Section 7 discusses defense mechanisms and limitations. Section 8 reviews related work, and Section 9 concludes the paper.
\section{Background}
\subsection{Bluetooth}

Bluetooth is a short-range wireless communication technology widely used by billions of personal computing, IoT, peripheral, and wearable devices for low-power communication. The technology operates in the 2.4GHz ISM band and enables devices to exchange commands and data such as keyboard/mouse inputs, audio, and files through secure communication channels.

The Bluetooth technology consists of two main variants: Bluetooth Basic Rate/Extended Data Rate (BR/EDR), also known as Bluetooth Classic, and Bluetooth Low Energy (BLE)\cite{puc2018ssp}. This work focuses on Bluetooth BR/EDR, which we refer to simply as Bluetooth throughout this paper. Bluetooth uses frequency hopping spread spectrum across 79 channels, each with a bandwidth of 1 MHz.

The Bluetooth system architecture comprises two main components: the Bluetooth Controller and the Bluetooth Host\cite{usenix2022braktooth}. The Controller implements the physical and logical layers in the Bluetooth chip, while the Host implements the L2CAP layer and application-oriented protocols in the device operating system. These components communicate through the Host Controller Interface (HCI).

In Bluetooth networks, devices are organized in a piconet structure where one device acts as the Master and up to seven other devices serve as Slaves. The Master provides the reference clock signal to synchronize all devices in the network. Importantly, device roles are not fixed—any Bluetooth device can initiate a connection and become the Master, regardless of its functionality or previous role. For example, Bluetooth headsets, typically slave devices, can take master roles and proactively initiate connections with smartphones. Devices can also switch Master-Slave roles after establishing a piconet.

\subsection{Bluetooth Pairing And Connection}
Establishing communication between Bluetooth devices involves two distinct processes: pairing and connection establishment\cite{dsn2022blap}. These processes implement multiple security mechanisms including encryption, authentication, and authorization to protect sensitive data transmission.

\subsubsection{Pairing Process}
Pairing is performed when two Bluetooth devices meet for the first time and need to establish a shared long-term key, known as the link key ($K_{L}$)\cite{ccs2024fake}. The most secure and widespread pairing mechanism is Secure Simple Pairing (SSP), which uses Elliptic Curve Diffie-Hellman (ECDH) for key agreement\cite{puc2018ssp,mdpi2019ssp,sp2022btdesign,sac2019curve}. If both devices support Secure Connections, SSP is performed on the P-256 curve; otherwise, it uses the P-192 curve\cite{sp2020bias}. During pairing, user confirmation is typically required when a new device attempts to pair with the user's device to prevent unauthorized connections.
The pairing process generates a link key that serves as the foundation for all future secure connections between the two devices. This key is stored on both devices and remains valid for subsequent connection attempts, eliminating the need to repeat the pairing process.

\subsubsection{Connection Establishment}
Once devices are paired and share a link key, they can establish multiple secure connections using different session keys derived from the long-term key and public parameters. However, the Bluetooth specification also supports temporary connections between unpaired devices for basic communication, service discovery, and legacy compatibility purposes.

The connection establishment process involves several phases:

\paragraph{Initial Connection Setup:} The Master device first queries the Slave to obtain device name and features, then sends a Connection Request to initiate connection establishment. This process leverages the flexible role assignment in Bluetooth, where the device initiating the connection becomes the Master.

\paragraph{Authentication}: Bluetooth provides two authentication mechanisms depending on device capabilities. Legacy Authentication is used for Legacy Secure Connections and provides unilateral authentication where typically only the Master authenticates the Slave\cite{ccs2023bluffs}. Secure Authentication is used for Secure Connections and provides mutual authentication where both devices must prove possession of the shared link key.

\paragraph{Encryption Key Negotiation}: After successful authentication, devices negotiate an encryption key for the session\cite{ndss2023pep}. The entropy of this encryption key ranges from 1 to 16 bytes according to the key length negotiation procedure. For Secure Connections, AES CCM encryption is used, while Legacy Secure Connections employ the $E_{0}$ stream cipher\cite{usenix2019knob,ccs2022blacktooth}.

\paragraph{Profile Management}: Bluetooth profiles define the protocols and functions required for specific device interactions\cite{ccs2022blacktooth}. Common profiles include the Advanced Audio Distribution Profile (A2DP) for audio communication, Human Interface Device Profile (HID) for keyboards and mice, and Phone Book Access Profile (PBAP) for contact access. Devices can implement multiple profiles simultaneously and advertise their available services using the Service Discovery Protocol (SDP).

\section{Threat Model}
\label{sec:threat_model}

\begin{figure}[t]
    \centering
    \includegraphics[width=\linewidth]{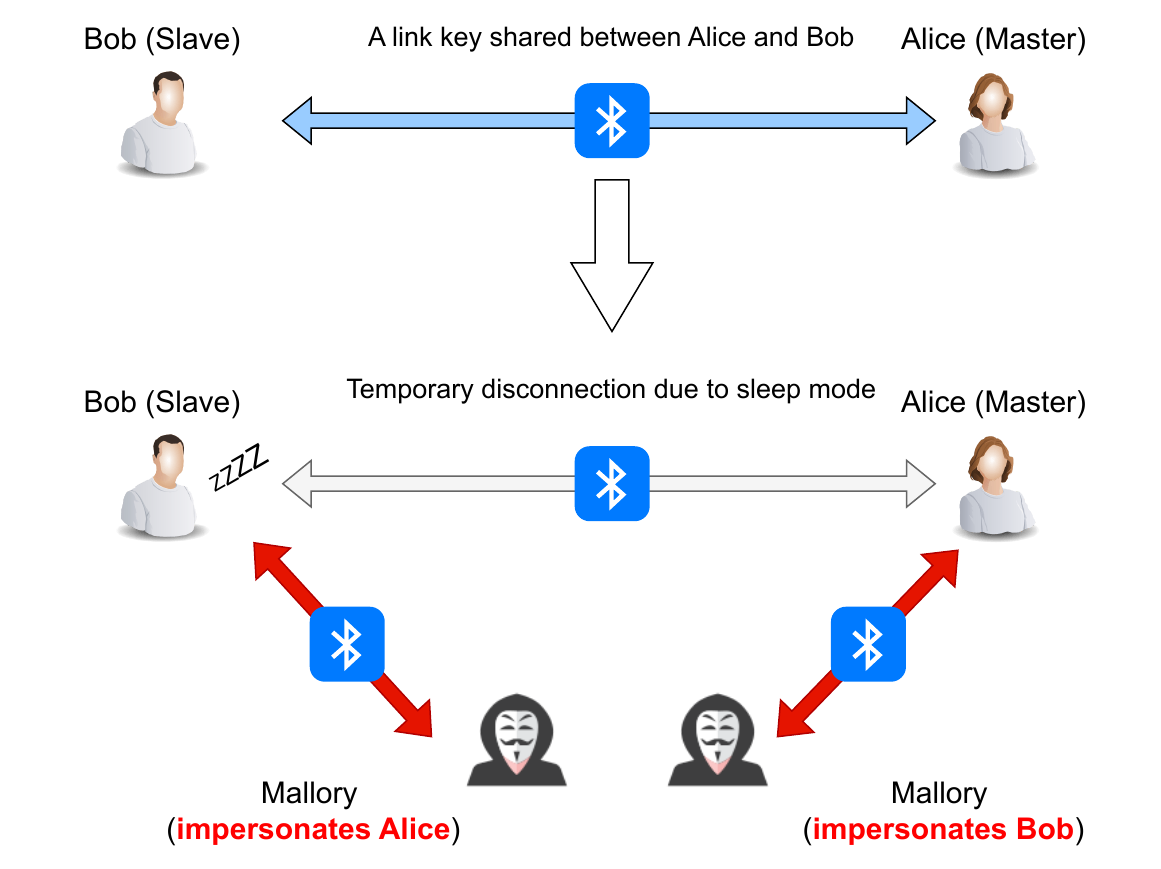}
    \caption[Stealtooth threat model]{Threat model: Alice and Bob share a link key in advance. Mallory does not know the link key pre-shared between the victims. Mallory aims to establish Bluetooth sessions with Alice and Bob and maliciously control communications and operations between Alice and Bob.}
    \label{fig:threat_model}
\end{figure}

In this section, we define our system and attacker models (Figure \ref{fig:threat_model}), as well as the notation we use in the rest of the paper.

\subsection{System Model}

We consider Alice and Bob (i.e., victims) who communicate securely via Bluetooth. Alice and Bob represent arbitrary Bluetooth devices and support arbitrary Bluetooth profiles (e.g., HID (Human Interface Device Profile), A2DP (Advanced Audio Distribution Profile), AVRCP (A/V Remote Control Profile)).

Without loss of generality, we assume Alice is the master and Bob is the slave. We assume the victims have previously paired and share a link key.

\subsection{Attacker Model}

We consider an attacker Mallory who aims to impersonate Alice, establish a secure connection with Bob, and use Bluetooth profiles to maliciously send data to Bob. Furthermore, Mallory also aims to impersonate Bob, establish a secure connection with Alice, and use the Bluetooth profile access permissions to control Alice's operations.

Mallory must be physically present within the Bluetooth range of the victims. Mallory does not know about the secure pairing process between the victims or their link keys. Mallory can capture unencrypted Bluetooth packets and recognize public information such as the victims' Bluetooth names and addresses.

\subsection{Notations}

In this paper, we denote the link key shared between Alice and Bob as $LK_{AB}$. Additionally, we denote the link key shared between Mallory and Alice as $LK_{MA}$, and the link key shared between Mallory and Bob as $LK_{MB}$.

\section{The Stealtooth Attack}
\label{sec:proposed_attack}

\subsection{Attack Description}

\subsubsection{Strategy}
\label{subsubsec:strategy}
The Stealtooth attack abuses a \textit{new} vulnerability that exists in Bluetooth's pairing mode, allowing Mallory to establish a \textit{completely stealthy} Bluetooth session with Bob. In this section, we explain the attack strategy of the Stealtooth attack.

The Stealtooth attack abuses a \textit{new} improper authentication vulnerability present in Bluetooth audio devices (i.e., Bob). The sequence of the vulnerability is shown in Figure \ref{fig:vuln_pattern_1} and Figure \ref{fig:vuln_pattern_2}. The vulnerability is such that when Mallory is impersonating a device paired with Bob (i.e., Alice), even if Bob is \textit{not} explicitly in pairing mode and \textit{without} requiring any operations from Bob's user, Bob will \textit{actively} (or passively, depending on the device) connect and pair with Mallory.
The attack strategy using the new vulnerability is as follows. First, Mallory changes her own Bluetooth address and adapter name to match Alice's settings. Next, Mallory sets the status of her Bluetooth adapter to \texttt{Discoverable} so that it can be detected by third-party devices\cite{sp2023bluesclues}. After setting the adapter status, Mallory waits until Bob has been inactive for a certain period of time or is manually powered off. Subsequently, when Alice is powered off or her Bluetooth adapter status is set to off, and Bob is powered on again, Bob will \textit{actively} (or passively, depending on the device) connect and pair with Mallory. At this time, during the pairing process, there are \textit{no} notifications or permission requests to the users of Bob or Alice, thereby achieving a \textit{completely stealth} attack.

\subsubsection{Reproducibility}
The Stealtooth attack is highly reproducible. To demonstrate that the attack is realistic, we describe a reproduction scenario. In the reproduction scenario, we consider a smartphone, Trent, which supports Bluetooth. Also, we assume that Alice is a laptop supporting Bluetooth, and Bob has already paired with both Alice and Trent. For example, Alice and Bob establish a Bluetooth session for a video conference. After the video conference ends, Alice and Bob turn off their power. Next, Trent attempts to connect with Bob to listen to music playing on the smartphone. However, if Mallory impersonating Alice exists within Bob's Bluetooth communication range, Bob improperly authenticates with Mallory before receiving the connection request from Trent. This reproduction scenario is the scenario in which we discovered the new vulnerability.

\subsubsection{Impact}
\label{subsubsec:impact}
The Stealtooth attack has a severe impact on Bluetooth security and privacy. The Stealtooth attack can easily overwrite the link key required for Bluetooth communication and maliciously hijack Bluetooth sessions between victims. Furthermore, using the hijacked session, Mallory can decrypt data transmitted from victims or inject authorized messages to victims. Importantly, the attack has \textit{high stealthiness} because pairing is executed \textit{without} triggering any user operations or notifications, despite the fact that Mallory has \textit{never} previously paired with Bob, meaning Mallory holds \textit{no} information about Bob whatsoever. Additionally, the attack strategy used by the Stealtooth attack is threatening because it is \textit{completely unaffected} by the presence or absence of SSP settings during Alice's or Bob's Bluetooth session establishment. Regarding the reproduction of the vulnerability used in the Stealtooth attack, it requires \textit{neither} special privileges or tools, \textit{nor} the implementation and execution of any specific code.

\subsection{Attack Root Causes}
\label{subsec:attacks_root_causes}

The root cause of the Stealtooth attack lies in the automatic pairing mode functionality implemented in Bob, which transitions to pairing mode without requiring user intervention. Hereafter, this functionality will be referred to as ``\textit{automatic pairing mode}''. Bob's automatic pairing mode is triggered when Bob attempts to reconnect with Alice but cannot confirm the connection. Mallory abuses this transition to automatic pairing mode to \textit{stealthily} overwrite the link key previously shared between Alice and Bob, circumventing both user intervention and notifications. The Stealtooth attack represents the \textit{first} attack that stealthily overwrites link keys by abusing the vulnerability in automatic pairing mode.

The link key overwrite scenario due to automatic pairing mode vulnerabilities can be divided into the following two patterns (\textbfit{\seqsplit{Pattern\#1}} and \textbfit{\#2}) based on the difference in connection initiators for overwriting the link key.

\paragraph{\textbfit{Pattern\#1: Bob is the connection initiator}}

\begin{figure}[t]
    \centering
    \includegraphics[width=\linewidth]{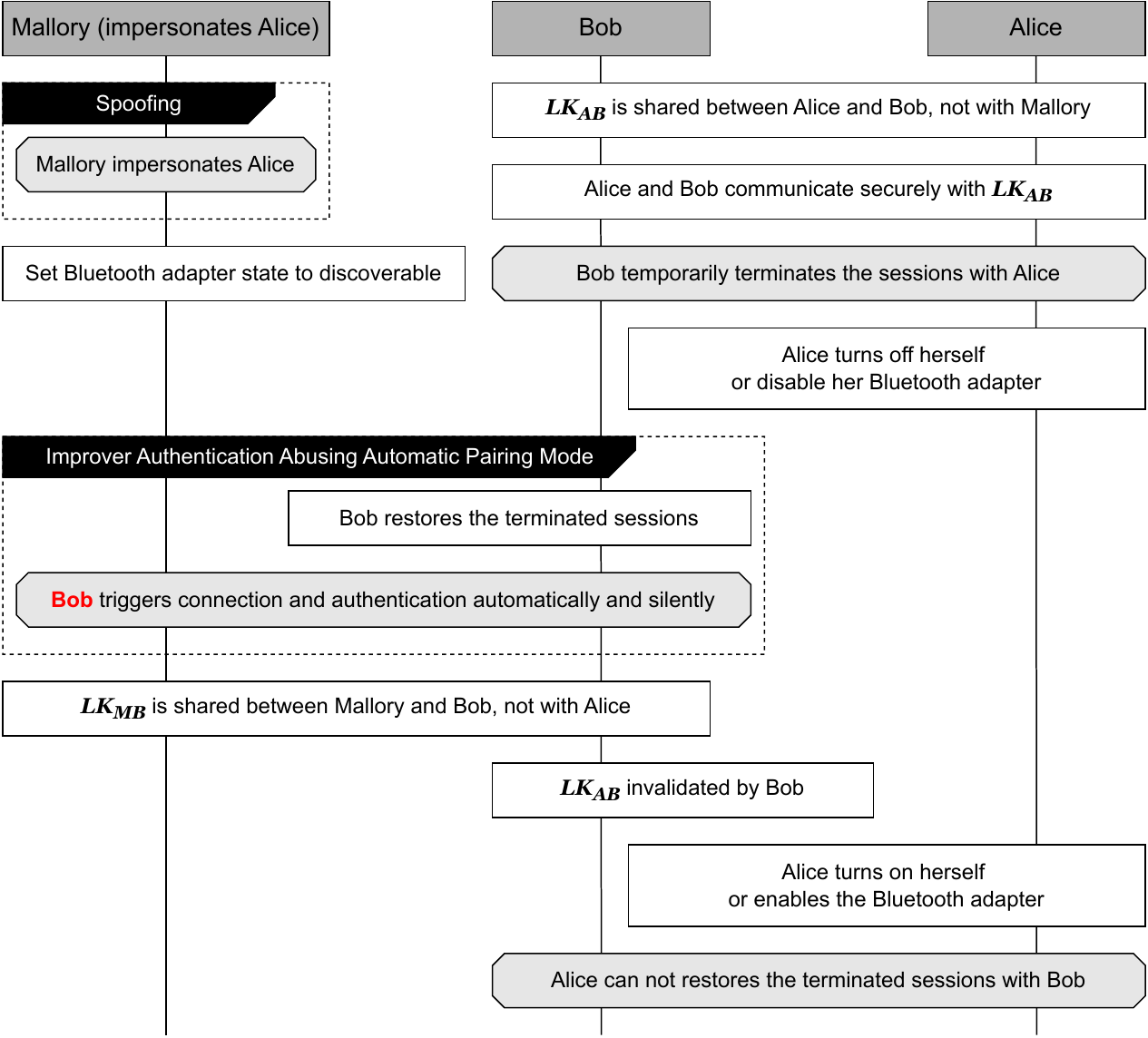}
    \caption[Improper Authentication Sequences (Bob is the connection initiator)]{Improper Authentication Sequences (Bob is the connection initiator): We assume that Alice and Bob are paired, and Mallory impersonates Alice and is discoverable to other devices. When Bob's Bluetooth session with Alice is disconnected and Alice is also unavailable for connection, Bob, upon being powered on again, attempts to connect to Mallory impersonating Alice and \textit{actively} executes pairing with Mallory through automatic pairing mode. To establish a malicious Bluetooth session with Bob, Mallory \textit{only} needs to impersonate Alice and wait for Bob's recovery.}
    \label{fig:vuln_pattern_1}
\end{figure}

Figure \ref{fig:vuln_pattern_1} shows the vulnerability sequence of improper authentication due to automatic pairing mode when the connection initiator is Bob. We assume that Alice and Bob are pre-paired and communicate securely. Meanwhile, Mallory impersonates Alice and makes herself discoverable to other Bluetooth devices. We consider the case where Bob's Bluetooth session with Alice is disconnected, and Alice is also unavailable for connection. In this scenario, when Bob recovers and restores the Bluetooth session, Bob attempts to connect to Mallory impersonating Alice. Since Bob misidentifies Mallory as Alice, device authentication fails due to the link key mismatch between Bob and Mallory. However, \textit{regardless} of the device authentication result, Bob \textit{actively} executes pairing while still misidentifying Mallory as Alice through automatic pairing mode. The connection initiator depends on the type of Bob device - the initiator may be either Bob or Mallory.

\paragraph{\textbfit{Pattern\#2: Mallory is the connection initiator}}

\begin{figure}[t]
    \centering
    \includegraphics[width=\linewidth]{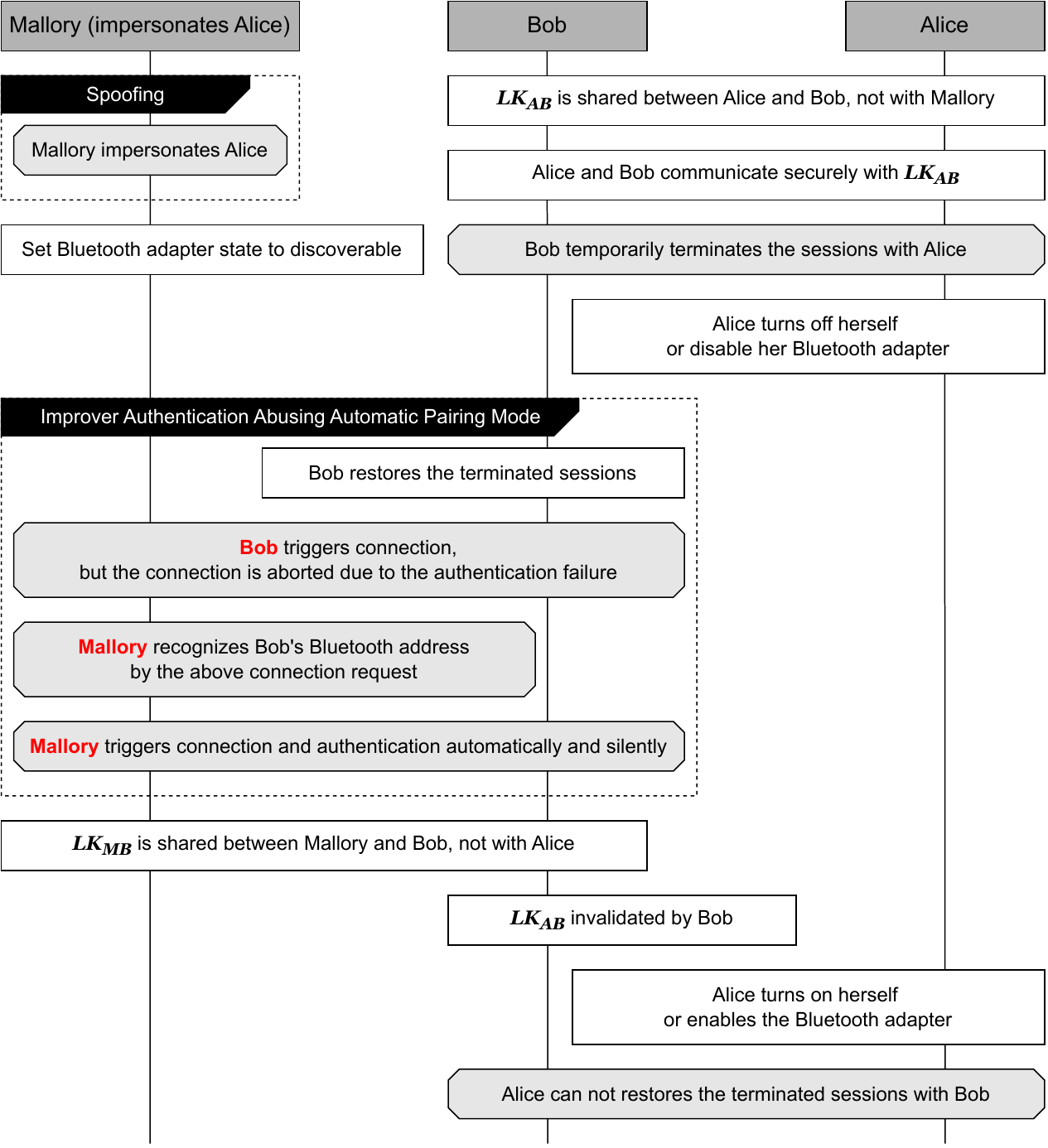}
    \caption[Improper Authentication Sequences (Mallory is the connection initiator)]{Improper Authentication Sequences (Mallory is the connection initiator): Considering the case where Bob is powered on again under the same assumptions as Figure \ref{fig:vuln_pattern_1}, Bob attempts to connect to Mallory impersonating Alice, but disconnects from Mallory due to device authentication failure caused by link key mismatch. However, through Bob's connection attempt, Mallory recognizes Bob's Bluetooth address. At this point, when Mallory requests a connection to Bob while still impersonating Alice, Bob in automatic pairing mode pairs with Mallory.}
    \label{fig:vuln_pattern_2}
\end{figure}

Figure \ref{fig:vuln_pattern_2} shows the vulnerability sequence of improper authentication due to automatic pairing mode when the connection initiator is Mallory. Under the same assumptions as \textbfit{Pattern\#1}, we consider the case where Bob attempts to restore the Bluetooth session again. Bob tries to connect to Mallory impersonating Alice. However, Mallory does not know the link key shared between Alice and Bob. Therefore, due to device authentication failure caused by the link key mismatch between Bob and Mallory, Bob disconnects the temporary connection with Mallory and transitions to automatic pairing mode. Meanwhile, through Bob's active connection attempt, Mallory recognizes Bob's Bluetooth address. At this point, when Mallory requests a connection to Bob while still impersonating Alice, Bob in automatic pairing mode pairs with Mallory regardless of the device authentication result and without requiring security-level verification during pairing.

The malicious establishment of Bluetooth sessions and link key overwriting through the above two patterns (\textbfit{\seqsplit{Pattern\#1}} and \textbfit{\#2}) both abuse the automatic pairing mode that devices secretly transition to. To the best of our knowledge, there have been \textit{no} prior attacks against Bluetooth that focused on the vulnerabilities of automatic pairing mode.

\subsection{The MitM Stealtooth Attacks}

While the Stealtooth attack is completely stealthy, victims may still have a chance to notice they are under attack. For example, if an attacker launches the Stealtooth attack when the victim wants to use their Bluetooth headset, the victim will find they cannot connect to their headset and have an opportunity to discover the attack. To make the attack more concealed, we extend the Stealtooth attack to a MitM attack.

To extend the Stealtooth attack to a MitM attack, we employ the Breaktooth attack proposed by Kimura et al. in 2025, which hijacks device operations using Bluetooth power-saving mode. In this section, we first provide an overview of the Breaktooth attack. Then, we describe the attack strategy of the MitM Stealtooth attack, which combines the Stealtooth and Breaktooth attacks.

\subsubsection{The Breaktooth Attack}

In 2025, Kimura et al. proposed Breaktooth, a device hijacking attack that abuses Sleep mode, a Bluetooth power-saving mode\cite{cryptoeprint2024breaktooth}. The Breaktooth attack is the first attack to hijack victim device operations using Sleep mode. The Breaktooth attack abuses two vulnerabilities in Sleep mode: \textbfit{Vuln.\#1}: the lack of security notifications when Bluetooth sessions are disconnected, and \textbfit{Vuln.\#2}: the vulnerability where the Master transitions to a state that accepts connection requests from Slaves after Bluetooth session disconnection. The attacker abuses \textbfit{Vuln.\#1} and \textbfit{\#2} to hijack Bluetooth sessions between victims without requiring any prior knowledge of link keys between victims, special privileges, or specialized tools. Furthermore, using the hijacked session as a starting point, the attacker gains complete control over the victim device operations.

The attack strategy of the Breaktooth attack consists of the following four steps (\textbfit{Step\#1-1} to \textbfit{\#1-4}):

\begin{description}
    \item[\textbfit{Step\#1-1. Spoofing}] The attacker changes their Bluetooth name and address to impersonate the victim's Slave.
    \item[\textbfit{Step\#1-2. Session Hijacking}] The attacker, impersonating the victim's Slave from \textbfit{Step\#1-1}, detects the temporary disconnection state of Bluetooth between victims due to Sleep mode, and at this moment sends a connection request to the Master as the victim's Slave to hijack the Bluetooth session between victims.
    \item[\textbfit{Step\#1-3. Link Key Hijacking}] After \textbfit{Step\#1-2}, the attacker, while still impersonating the victim's Slave, sends a pairing request to the Master specifying low security-level authentication functions and generates a new link key between them.
    \item[\textbfit{Step\#1-4. Command Injection}] Exploiting the link key hijacked in \textbfit{Step\#1-3}, the attacker uses sensitive profiles to control the victim Master's operations. For example, the attacker uses the Human Interface Device profile to inject malicious commands into the victim's Master.
\end{description}

\begin{figure}[t]
    \centering
    \includegraphics[width=\linewidth]{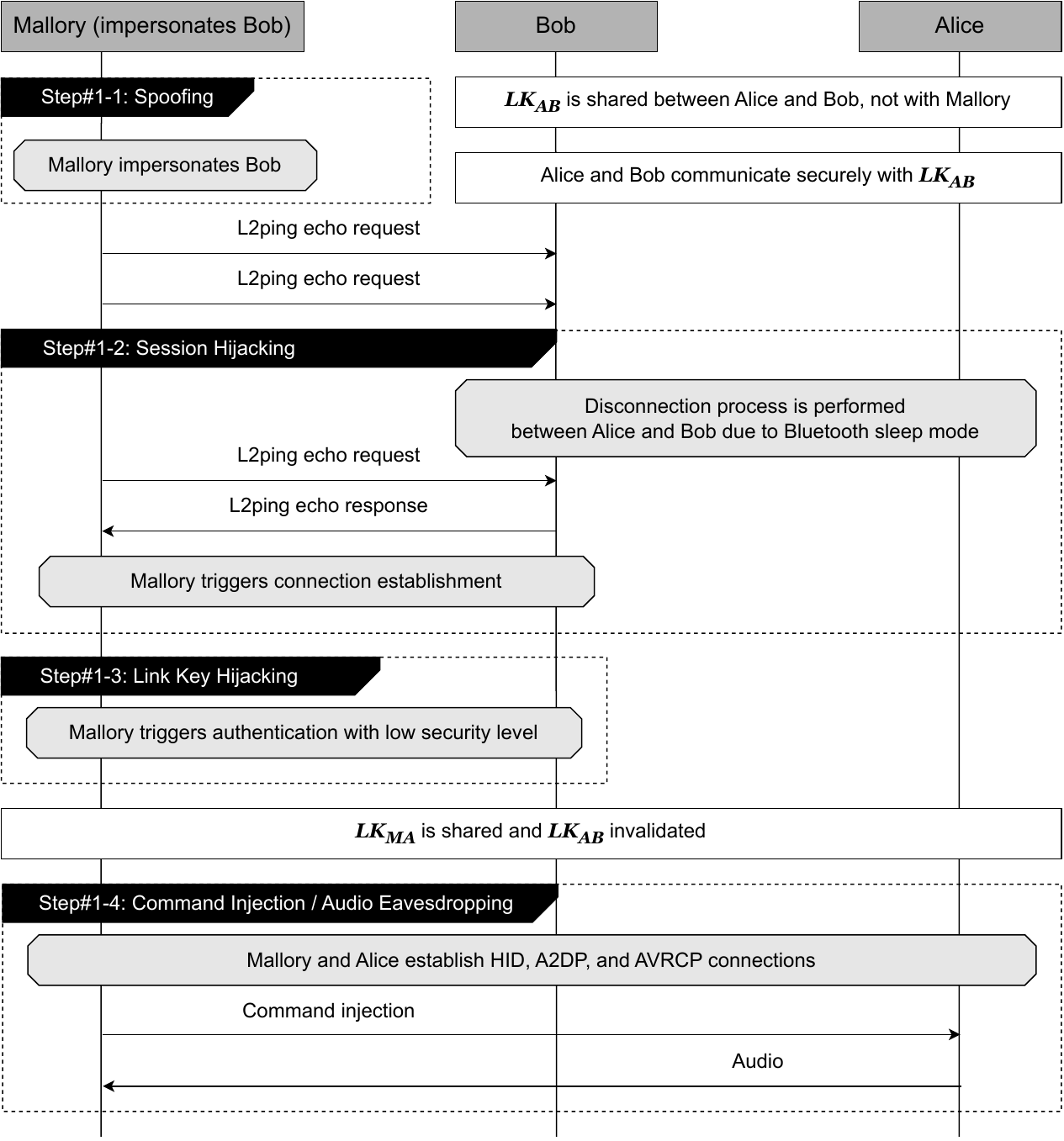}
    \caption[The Breaktooth Attack]{The Breaktooth attack sequences: The Breaktooth attack abuses a vulnerability where Alice transitions to a state that accepts reconnection requests from Bob after Bob enters Sleep mode as the attack starting point. Mallory detects Sleep mode from Alice's response behavior to malicious \texttt{l2ping} echo requests and abuses the Sleep mode vulnerability to hijack the Bluetooth session between Alice and Bob.}
    \label{fig:breaktooth}
\end{figure}

The sequence of the Breaktooth attack is shown in Figure \ref{fig:breaktooth}. To hijack sessions by abusing Sleep mode vulnerabilities, the attacker needs to remotely and secretly monitor the state of Bluetooth sessions between victims. The Breaktooth attack achieves this by maliciously using \texttt{l2ping}\cite{ssrn2022l2ping,ccs2024bluewswat,secrypt2024bluedos,l2ping}. Kimura et al. analyzed the echo request and response behavior of \texttt{l2ping}, enabling attackers to remotely and secretly recognize Bluetooth sessions between victims.

Kimura et al. developed a tool to demonstrate the Breaktooth attack and evaluated the attack against commercial Bluetooth keyboards, mice, and audio devices that support Sleep mode. The evaluation results showed that attackers could control the operations of Masters (e.g., laptops, smartphones). Additionally, Kimura et al. released the developed tool as open-source.

\subsubsection{MitM Attacks Combining Stealtooth And Breaktooth}
\label{subsubsec:mitm_blindtooth}

\begin{figure*}[t]
    \centering
    \includegraphics[width=\linewidth]{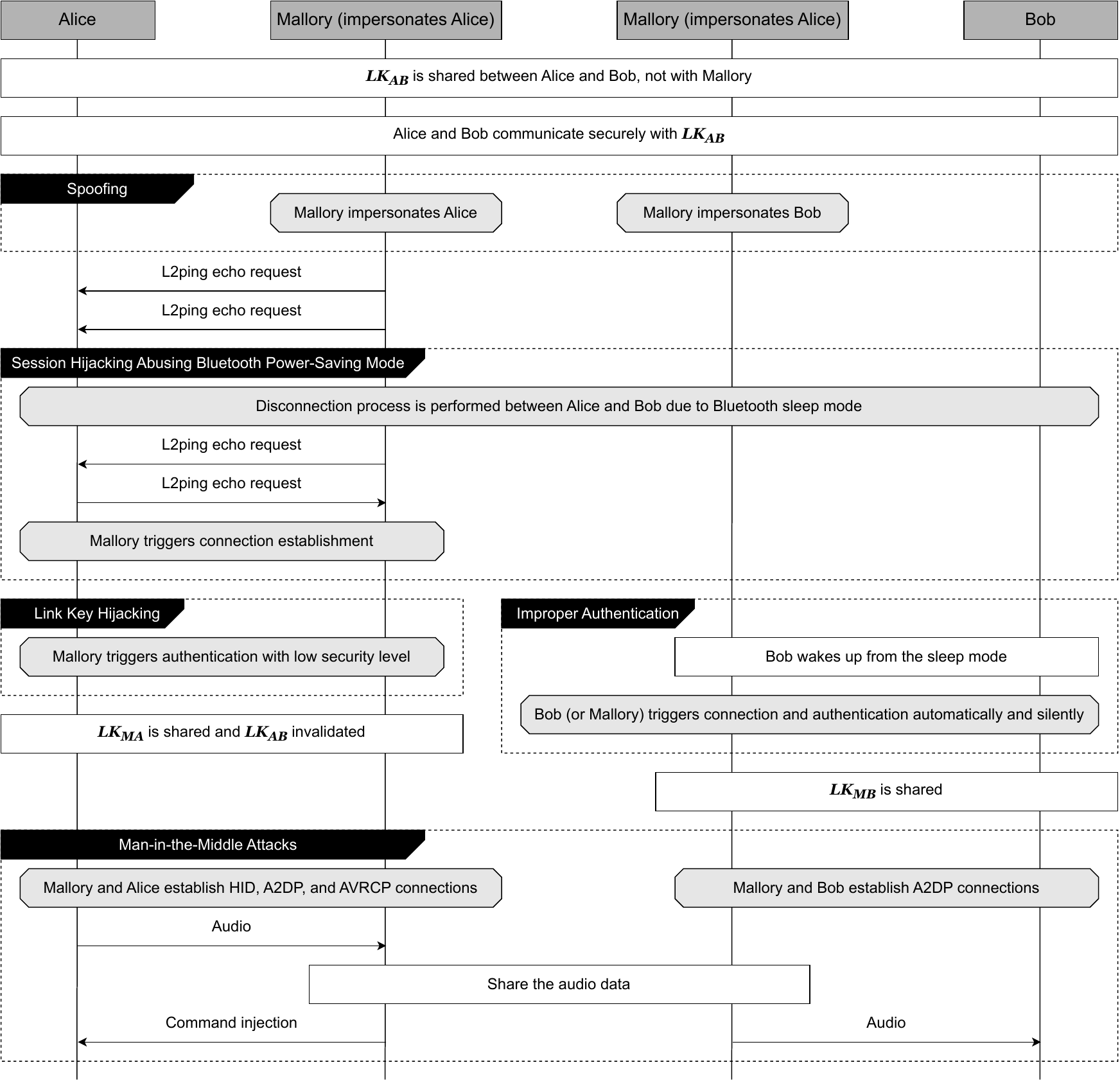}
    \caption[MitM]{The MitM Stealtooth attack strategy: Alice and Bob have already paired and share $LK_{AB}$. If Bob remains inactive for a certain period, Bob terminates the session between Alice and Bob. After the termination, Mallory triggers connection establishment and low-level authentication. After the authentication, Mallory and Alice share $LK_{MA}$. On the other hand, when Bob restores the session, Mallory abuses the authentication vulnerability to pair with Bob, and shares $LK_{MB}$ with Bob. Finally, Mallory stealthily proxies the communication between Alice and Bob as a MitM abusing $LK_{MA}$ and $LK_{MB}$.}
    \label{fig:mitm}
\end{figure*}

This section introduces the MitM Stealtooth attack strategy. 
Figure \ref{fig:mitm} shows the details of the attack strategy. 
The strategy consists of the following five steps (\textbfit{Step\#2-1} to \textbfit{\#2-5}):

\begin{description}
    \item[\textbfit{Step\#2-1. Spoofing}] Mallory changes her Bluetooth name and its address to impersonate Bob. The same applies when Mallory impersonates Alice. 
    \item[\textbfit{Step\#2-2. Session Hijack}] Mallory, impersonating Bob, detects the temporary disconnection state of Bluetooth between Alice and Bob by the sleep mode, and at this moment sends a connection request to Alice as Bob, hijacking the Bluetooth session between Alice and Bob.
    \item[\textbfit{Step\#2-3. Link Key Hijack}] After the session hijack attack, Mallory sends a pairing request to Alice while impersonating Bob, generating a new link key between them ($LK_{MA}$). This invalidates $LK_{AB}$. Mallory does this by bypassing the PIN code authentication.
    \item[\textbfit{Step\#2-4. Improper Authentication}] Mallory, impersonating Alice, abuses the authentication vulnerability described in Section \ref{subsubsec:strategy} and actively triggers Bob, who has returned from the sleep mode, to perform authentication, generating and sharing $LK_{MB}$ between Mallory and Bob.
    \item[\textbfit{Step\#2-5. MitM Attacks}] Up to the improper authentication step, Mallory shares the link key $LK_{MA}$ with Alice and the link key $LK_{MB}$ with Bob. Therefore, Mallory becomes a MitM, capable of intercepting and manipulating the communication between Alice and Bob. For example, Mallory, impersonating Bob, establishes a connection with Alice using both the A2DP and AVRCP profiles, while Mallory, impersonating Alice, establishes a connection with Bob using the A2DP profile. In this scenario, Mallory can secretly eavesdrop on the audio data transmitted between Alice and Bob.
\end{description}

\section{Implementation}
\label{sec:impl}

As described in Section \ref{subsubsec:impact}, the execution of the Stealtooth attack requires no specific attack code whatsoever. Therefore, this chapter focuses on the implementation of the MitM Stealtooth attack.

As described in Section \ref{subsubsec:mitm_blindtooth}, among the attack strategies of the MitM attack, \textbfit{Step\#2-1} to \textbfit{Step\#2-3} are identical to the Breaktooth attack. Therefore, we execute them using the open-source tool of the Breaktooth attack. Additionally, \textbfit{Step\#2-4} can be executed without requiring implementation. Therefore, this chapter describes the implementation details for demonstrating the man-in-the-middle attack in \textbfit{Step\#2-5}.

We implement the following two systems to intercept and relay audio data transmitted and received using A2DP between victim Bluetooth devices.

\begin{description}
    \item[\textbf{A2DP Sender}:] A2DP Sender operates on Mallory impersonating Bob. A2DP Sender forwards audio data received from Alice to the other Mallory impersonating Alice.
    \item[\textbf{A2DP Receiver}:] A2DP Receiver operates on Mallory impersonating Alice. A2DP Receiver sends audio data forwarded from A2DP Sender to Bob.
\end{description}

\paragraph{A2DP Sender Implementation}
\label{subsec:a2dp_sender}

A2DP Sender intercepts A2DP streams transmitted from Alice and forwards them to A2DP Receiver via TCP socket communication. For obtaining A2DP streams, we utilize the Bluetooth module of PulseAudio server. PulseAudio server cooperates with the A2DP stack provided by BlueZ to acquire Bluetooth audio data as PCM data.

To ensure session stability, we implement an independent thread that monitors TCP socket status and attempts reconnection with exponential backoff when disconnection is detected. We also manage captured PCM data with queue buffers to mitigate the effects of network delays.

\paragraph{A2DP Receiver Implementation}
\label{subsec:a2dp_receiver}

A2DP Receiver sends audio data forwarded from A2DP Sender to Bob using A2DP. This system operates as a TCP socket server and utilizes the BlueZ stack through PulseAudio server for audio stream transmission. Using PulseAudio enables retransmission of intercepted audio data as A2DP-compliant audio streams. Additionally, by leveraging the audio processing stack of PulseAudio and BlueZ, we implement conversion from PCM data to audio codecs used in A2DP (e.g., SBC) and stable audio stream transmission.

Our implementation assumes the BlueZ stack on the Linux kernel. We adopted BlueZ for attack demonstration because it is the most widely adopted open-source Bluetooth protocol stack.

\section{Evaluation}
\label{sec:eval}

In this section, we describe devices used for the Stealtooth attack evaluation and attack scenarios to evaluate the attack. We also present our evaluation setup and results.

\subsection{Attack Devices}

In the Stealtooth attack evaluation, we use a Raspberry Pi 4 Model B device as Mallory, 
as shown in Table \ref{tab:devices_as_mallory}. 
In the MitM Stealtooth attack evaluation, we use two Raspberry Pi shown in Table \ref{tab:devices_as_mallory}.

The Raspberry Pi runs on Raspberry Pi OS (11 Bullseye) with Linux OS kernel version 6.1. This operating system comes with BlueZ 5.55 pre-installed. We employ \texttt{hciconfig} for HCI configuration commands and \texttt{hcitool} for Bluetooth device scanning and enumeration operations\cite{springer2025hcitool}. Both commands are included in the BlueZ Bluetooth stack and are readily available upon OS installation. Additionally, Bluetooth adapter support is built into the Raspberry Pi by default. Consequently, the hardware/software cost of the Stealtooth attack is \textit{low}.

Additionally, we use one device as Alice, as shown in Table \ref{tab:device_of_alice}. 
For Bob, we use 10 commercial Bluetooth headsets as shown in Table \ref{tab:devices_of_bob}. 

\begin{table}[t]
    \centering
    \caption[Specifications of devices used as Mallory in the Stealtooth attack]
    {\textbf{Specifications of devices used as Mallory in the Stealtooth attack}: 
    In case of the MitM Stealtooth attack evaluation, two devices are required as Mallory - one impersonating Alice and one impersonating Bob, but both use devices with the same specifications.}
    \label{tab:devices_as_mallory}
    \footnotesize
    \begin{tabularx}{\linewidth}{>{\centering\arraybackslash}X>{\centering\arraybackslash}X}
         \toprule
         Device Model      & Raspberry Pi 4 Model B \\
         Operating System  & Raspberry Pi OS        \\
         System            & 32bit                  \\
         Debian Version    & 11 Bullseye            \\
         Kernel Version    & 6.1                    \\
         BlueZ Version     & 5.55                   \\
         Bluetooth Version & 5.0                    \\
         \hline
    \end{tabularx}
\end{table}

\begin{table}[t]
    \centering
    \caption[Specifications of device used as Alice in the Breaktooth attack]
    {\textbf{Specifications of device used as Alice in the Breaktooth attack}}
    \label{tab:device_of_alice}
    \footnotesize
    \begin{tabular}{cccc}
        \toprule
        Manufacturer   & Model & Operating System & Bluetooth Version \\
        \midrule
        Microsoft & Surface Laptop 4 & Windows 11 & 5.1 \\
        \bottomrule
    \end{tabular}
\end{table}

\begin{table*}[t]
    \centering
    \caption{\textbf{Specifications of devices used as Bob}}
    \label{tab:devices_of_bob}
    \small
    \begin{tabular}{ccccc}
        \toprule
        Manufacturer   & Model & Bluetooth Version & Chip Producer & Chip Model \\
        \midrule
        Sony    & WH-1000XM5                         & 5.2 & MediaTek & MT2822 \\
        Sony    & WH-1000XM4                         & 5.0 & MediaTek & MT2811 \\
        Sony    & WF-1000XM5                         & 5.3 & - & - \\
        Sony    & WF-1000XM4                         & 5.2 & MediaTek & MT2822S \\
        Anker   & Soundcore Space One                & 5.3 & - & - \\
        EDIFIER & W820NB                             & 5.0 & - & - \\
        TOZO    & NC2                                & 5.2 & - & - \\
        Xaomi   & Redmi Buds 6 Pro                   & 5.3 & - & - \\
        Google  & Pixel Buds Pro                     & 5.0 & - & -  \\ 
        BOSE    & Bose QuietComfort Ultra Headphones & 5.3 & Qualcomm & QCC5181 \\ 
        \bottomrule
    \end{tabular}
\end{table*}

\subsection{Attack Scenarios}

To evaluate the effectiveness of the MitM Stealtooth attack as a man-in-the-middle attack, we define the following three attack scenarios: \textbfit{AS\#1, \#2, \#3}.

\begin{description}
    \item[\textbfit{AS\#1. Interception of communication}:] Mallory attempts to eavesdrop on audio data transmitted from Alice to Bob.
    \item[\textbfit{AS\#2. Tampering with communication}:] Mallory evaluates whether it is possible to send different audio data to Bob instead of the audio data received from Alice.
    \item[\textbfit{AS\#3. Proxying communication}:] Mallory evaluates whether audio data received from Alice can be seamlessly forwarded to Bob. 
\end{description}

\subsection{Setup}

\begin{figure}[t]
    \centering
    \includegraphics[width=\linewidth]{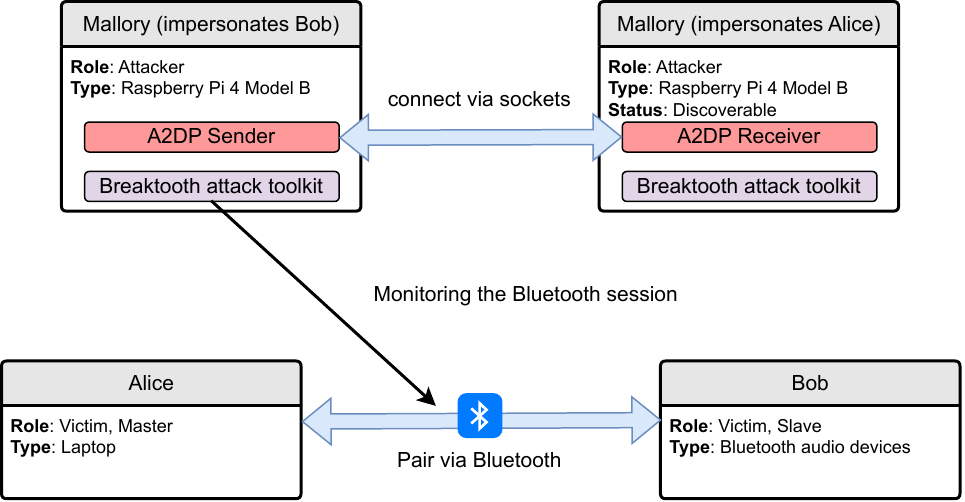}
    \caption[Evaluation model for the MitM Stealtooth attack]{Evaluation model for the MitM Stealtooth attack: Alice and Bob are paired in advance. Meanwhile, both Raspberry Pis operating as Mallory have installed the Breaktooth attack toolkit. Additionally, A2DP Sender and A2DP Receiver implemented in \textbf{Section \ref{sec:impl}} are launched, and communication is established between the two Mallory devices in advance}
    \label{fig:mitm_eval_model}
\end{figure}

To evaluate the Stealtooth attack, we first pair the victims Alice and Bob and establish a Bluetooth session. Next, the attacker Mallory sets one device's Bluetooth name and address to match those of Alice.

Figure \ref{fig:mitm_eval_model} shows the evaluation model of the MitM Stealtooth attack. For the evaluation of the MitM attack, after pairing the victims Alice and Bob and establishing a Bluetooth session, we keep Bob in an inactive state for a certain period until Bob enters idle mode. The attacker Mallory sets one device's Bluetooth name and address to match those of Bob, and the other to match those of Alice. We then install the attack tool ``Breaktooth'' on both Mallory devices. The Mallory impersonating Alice remotely and secretly monitors the Bluetooth session state between Alice and Bob until Bob transitions to idle state. The Bluetooth session state can be monitored using Breaktooth's functionality. Additionally, to transmit audio data intercepted from Alice to the other Mallory, we launch the A2DP Sender implemented in Section \ref{subsec:a2dp_sender}. Meanwhile, the Mallory impersonating Bob sets its Bluetooth adapter's \texttt{Discoverable} parameter to \texttt{yes}, making the Bluetooth adapter detectable. Furthermore, to receive audio data transmitted from the A2DP Sender, we launch the A2DP Receiver implemented in Section \ref{subsec:a2dp_receiver}.

\subsection{Results}

\begin{table*}[t]
    \centering
    \caption{\textbf{Stealtooth attack evaluation results}}
    \label{tab:results}
    \small
    \begin{tabular}{cccccc}
        \toprule
        \multicolumn{1}{c}{Model (Bob)} & \multicolumn{2}{c}{Stealtooth Attack} & \multicolumn{3}{c}{MitM Stealtooth Attack} \\
        \cmidrule(lr){2-3} \cmidrule(lr){4-6}
        & Attack & Connection Initiator & \textbfit{AS\#1} & \textbfit{AS\#2} & \textbfit{AS\#3} \\
        \midrule
         WH-1000XM5                             & \textcolor{red}{$\tikzcmark$} & Bob        & $\fullcirc$ & $\fullcirc$ & $\halfcirc$ \\
         WH-1000XM4                             & \textcolor{red}{$\tikzcmark$} & Mallory    & $\fullcirc$ & $\halfcirc$ & $\halfcirc$ \\
         WF-1000XM5                             & \textcolor{red}{$\tikzcmark$} & Bob        & $\fullcirc$ & $\fullcirc$ & $\halfcirc$ \\
         WF-1000XM4                             & \textcolor{red}{$\tikzcmark$} & Mallory    & $\fullcirc$ & $\halfcirc$ & $\halfcirc$ \\
         Soundcore Space One                    & \textcolor{red}{$\tikzcmark$} & Mallory    & - & - & - \\
         W820NB                                 & \textcolor{red}{$\tikzcmark$} & Mallory    & - & - & - \\
         NC2                                    & \textcolor{red}{$\tikzcmark$} & Mallory    & - & - & - \\
         Redmi Buds 6 Pro                       & \textcolor{red}{$\tikzcmark$} & Mallory    & - & - & - \\
         Pixel Buds Pro                         & \textcolor{blue}{$\tikzxmark$} & -          & - & - & - \\
         Bose QuietComfort Ultra Headphones     & \textcolor{blue}{$\tikzxmark$} & -          & - & - & - \\
        \bottomrule
    \end{tabular}
    \begin{itemize}
        \footnotesize
        \item[\textcolor{red}{$\tikzcmark$}] The Stealtooth attack is achieved.
        \item[\textcolor{blue}{$\tikzxmark$}] The Stealtooth attack is not achieved.
        \item[$\fullcirc$] Attack scenario is achieved.
        \item[$\halfcirc$] Attack scenario is partially achieved.
        \item[$\emptycirc$] Attack scenario is not achieved.
    \end{itemize}
\end{table*}

In this section, we present the evaluation results of our Stealtooth and MitM Stealtooth attacks against commercial Bluetooth devices.

\subsubsection{Stealtooth Attack Results}
The Stealtooth attack evaluation results are shown in Table \ref{tab:results}. We successfully demonstrated the Stealtooth attack against 8 out of 10 tested devices, including Sony WH-1000XM5, WH-1000XM4, WF-1000XM5, WF-1000XM4, Anker Soundcore Space One, EDIFIER W820NB, and TOZO NC2. The attack was not successful against Google Pixel Buds Pro and BOSE QuietComfort Ultra Headphones.

The results reveal two distinct patterns based on the connection initiator. For Sony WH-1000XM5 and WF-1000XM5, Bob acts as the connection initiator (\textbfit{Pattern\#1}), actively attempting to reconnect with Alice and subsequently pairing with Mallory through automatic pairing mode. For the remaining vulnerable devices, Mallory serves as the connection initiator (\textbfit{Pattern\#2}), where Bob transitions to automatic pairing mode after failed authentication attempts, enabling Mallory to establish malicious connections.

Notably, the vulnerability appears to be implementation-specific rather than chip-specific, as devices with similar MediaTek chipsets (MT2822 and MT2822S) exhibit different behaviors. The successful attacks demonstrate complete link key overwriting without any user notification or intervention, confirming the silent nature of the automatic pairing mode abuse.

\subsubsection{MitM Stealtooth Attack Results}
The MitM Stealtooth attack results are also shown in Table \ref{tab:results}. Among the 8 devices vulnerable to the Stealtooth attack, 4 devices are also vulnerabile to the MitM Stealtooth attack.

For attack scenario \textbfit{AS\#1} (interception of communication), the MitM Stealtooth attack successfully demonstrated the ability to intercept audio data transmitted from Alice to Bob. The attack captured and recorded the intercepted audio as WAV files, confirming the compromised confidentiality of Bluetooth communications.

Attack scenario \textbfit{AS\#2} (tampering with communication) was achieved against Sony WH-1000XM5 and WF-1000XM5 devices. The attack successfully demonstrated that while Mallory intercepts audio data from Alice, Mallory can simultaneously play pre-prepared audio data to Bob, effectively tampering with the communication content. For Sony WH-1000XM4 and WF-1000XM4, where Mallory serves as the connection initiator, AS\#2 was achieved only after establishing the connection, disconnecting briefly, and reconnecting to properly grant profile access permissions.

Attack scenario \textbfit{AS\#3} (proxying communication) was partially achieved for the four tested devices. While Mallory impersonating Bob could transfer intercepted audio data from Alice via A2DP Sender, and Mallory impersonating Alice could receive the transferred audio data with A2DP Receiver, the received audio data could not be properly encoded into the appropriate audio codec when transmitting to Bob, preventing seamless audio playback.

\subsubsection{Impact Assessment}
The evaluation results demonstrate that abusing automatic pairing vulnerabilities enables attackers to become MitM adversaries in Bluetooth communications between victims. The successful achievement of attack scenarios \textbfit{AS\#1} and \textbfit{AS\#2} confirms that the MitM Stealtooth attack significantly compromises the confidentiality, integrity and availability of Bluetooth communications.

The widespread nature of these vulnerabilities across major manufacturers and different device types indicates that automatic pairing implementations represent systematic security weaknesses rather than isolated flaws. The fact that 80\% of tested devices were vulnerable to the Stealtooth attack, with 40\% were vulnerable to the MitM attacks, underscores the critical impact of these findings on the broader Bluetooth ecosystem.

The results also reveal that the success of automatic pairing exploitation depends heavily on device-specific implementations, suggesting that manufacturers have varying approaches to automatic pairing that create \textit{inconsistent} security behaviors across the market.
\section{Discussion}

In this section, we propose defenses against the Stealtooth attack. We also describe limitations of this work and future work. 

\subsection{Defense Against the Stealtooth Attack}
To defend against the Stealtooth attack, we propose several defenses that can be implemented at different levels of the Bluetooth stack.

\subsubsection{Device-Level Defenses}
The most immediate defense against the Stealtooth attack lies in improving the transparency and control of automatic pairing mode. Bluetooth devices should notify users when they automatically transition to pairing mode and when new pairings are established. This notification mechanism would alert users to potential unauthorized pairing attempts and allow them to take appropriate action. However, such notifications must be carefully designed to balance security awareness with user experience, as excessive notifications could lead to alert fatigue.

Manufacturers should also consider implementing additional authentication layers for automatic pairing scenarios. The authentication layers could include device-specific challenges or cryptographic proofs that prevent unauthorized devices from successfully completing automatic pairing. Such mechanisms would need to be carefully designed to avoid breaking compatibility with existing paired devices while preventing unauthorized pairings.

\subsubsection{Protocol-Level Improvements}
At the protocol level, the Bluetooth specification should include mechanisms to validate the legitimacy of link key replacement during automatic pairing scenarios. This could involve cryptographic proofs or time-based validation tokens that ensure only legitimate devices can replace existing pairings. Such improvements would require careful consideration of backward compatibility and the diverse ecosystem of Bluetooth implementations.

Furthermore, the Bluetooth SIG should establish clear guidelines for automatic pairing implementations to ensure consistent security behaviors across different manufacturers and device types. The current lack of standardization in automatic pairing behavior creates inconsistent attack surfaces and makes it difficult to develop comprehensive defensive strategies.

\subsection{Limitations and Future Work}

\subsubsection{Current Limitations}
Our current implementation and evaluation have several limitations that constrain the scope of our findings. While we evaluated 10 commercial Bluetooth devices from major manufacturers, the automatic pairing behaviors vary significantly across device types and firmware versions. The diversity of implementations means that our findings may not generalize to all Bluetooth devices in the market. We need a more comprehensive evaluation across broader device categories and manufacturers to understand the vulnerability landscape.

The success of the Stealtooth attack also depends on specific timing conditions, such as when legitimate devices are powered off or unavailable. In real-world scenarios, these conditions may not always be controllable by attackers, limiting the practical applicability of the attack. The probabilistic nature of these timing dependencies means that attacks may require multiple attempts to succeed.

Our MitM Stealtooth attack implementation focuses primarily on A2DP audio interception, representing only a subset of the potential capabilities enabled by successful device impersonation. Extension to other Bluetooth profiles and more sophisticated manipulation capabilities would require additional implementation work.

\subsubsection{Future Work}

Our work opens several important research directions that could significantly advance the understanding of Bluetooth security and automatic pairing vulnerabilities.
In this section, we present two main future directions of this work.

\paragraph{Cross-Platform Vulnerability Analysis.} While our evaluation focuses on commercial audio devices, automatic pairing modes exist across a much broader ecosystem of Bluetooth-enabled devices, including IoT sensors, medical devices, automotive systems, and industrial controllers. These diverse platforms may implement automatic pairing with different security assumptions and constraints, potentially exposing novel attack vectors. A systematic analysis of automatic pairing implementations across heterogeneous platforms could reveal fundamental design patterns that are inherently vulnerable, leading to more comprehensive mitigation strategies that address root causes rather than individual device vulnerabilities.

\paragraph{Large-Scale Empirical Security Assessment.} The true scope of automatic pairing vulnerabilities in deployed devices remains unknown. Conducting large-scale empirical studies that systematically test thousands of devices across different manufacturers, firmware versions, and deployment contexts would provide crucial data for understanding the real-world impact of these vulnerabilities. Such studies could also reveal correlations between device characteristics and vulnerability patterns, informing risk assessment and prioritization of mitigation efforts.

\section{Related Work}

In this section, we describe prior MitM attacks against Bluetooth and present the differences between these prior attacks and the MitM Stealtooth attack.

\subsection{Prior MitM Attacks Against Bluetooth}

\subsubsection{MitM Attack Abusing Unidirectionality of Authentication Procedure}
\label{subsubsec:about_bias}

The Bluetooth Impersonation AttackS (BIAS) proposed by Antonioli et al. is an attack that abuses the unidirectional vulnerability in Bluetooth's authentication procedure, allowing attackers to impersonate the victim's device and establish a connection\cite{sp2020bias}. Antonioli et al. evaluated the BIAS attack against 28 types of Bluetooth chips and demonstrated its effectiveness.

The BIAS attack can become a MitM attack when combined with the KNOB attack. Attackers can establish independent encrypted communication channels with both the sender and receiver devices using the BIAS attack, and then reduce the entropy of the session key using the KNOB attack to eavesdrop on or tamper with communications. The attack combining BIAS and KNOB attacks does not require any user operation or malicious applications. Furthermore, it can be executed on any Bluetooth device that complies with the standard Bluetooth protocol.

\subsubsection{MitM Attack Using the Variability of Master/Slave Roles}
\label{subsubsec:about_blactooth}

The Blacktooth attack proposed by Ai et al. is the first attack that exploits the specification that Bluetooth Master/Slave roles are not fixed, allowing attackers to covertly complete connection and pairing with the victim's device\cite{ccs2022blacktooth}. The attacker establishes a connection by impersonating the legitimate device and bypasses authentication by exploiting the unidirectionality of legacy authentication. Additionally, they decrypt communication contents using the aforementioned KNOB attack to gain access permissions to highly confidential profiles and inject unauthorized commands to the victim. The Blacktooth attack can be executed without requiring any user operation or malicious applications. Ai et al. demonstrated the effectiveness of the Blacktooth attack against 21 types of Bluetooth devices.

\subsubsection{MitM Attack Abusing Differences in Authentication Methods}
\label{subsubsec:about_mc}

The Method Confusion (MC) attack proposed by Tschirschnitz et al. is a MitM attack that exploits differences in authentication methods during Bluetooth pairing\cite{sp2021mc,ndss2025mc}. In the MC attack, the attacker completes pairing as a MitM by using Numeric Comparison authentication with one device and Passkey Entry authentication with the other device. Tschirschnitz et al. evaluated the effectiveness of the MC attack with 40 users using devices such as smartwatches and smartphones, and successfully executed the MC attack against 37 users.

\subsection{Comparison Between Prior MitM Attacks and the MitM Stealtooth Attack}

The most significant difference between the prior attacks described in Section 2.3 and the conventional attacks and proposed attack lies in whether Bluetooth jamming between victims is necessary to initiate the attack.

The BIAS attack described in Section \ref{subsubsec:about_bias} requires forcibly disconnecting the Bluetooth session between victim devices to execute the attack. Therefore, Antonioli et al. assumed that in the attacker model for the BIAS attack, the attacker can jam the Bluetooth spectrum and has the ability to forcibly disconnect Bluetooth. However, they did not describe specific methods or feasibility of jamming.

The Blacktooth attack described in Section \ref{subsubsec:about_blactooth}, like the BIAS attack, requires forcibly disconnecting the Bluetooth session between victim devices to execute the attack. However, similar to Antonioli et al., Ai et al. did not describe specific methods or feasibility of jamming.

The MC attack described in Section \ref{subsubsec:about_mc} also assumes jamming for its execution. The MC attack describes a specific method of selectively interfering with Bluetooth advertisement packets as ``selective jamming''. Unlike the BIAS attack and the Blacktooth attack, selective jamming does not jam the entire Bluetooth spectrum but only interferes with targeted packets. Selective jamming makes it possible to minimize the impact on other devices, making the attack less conspicuous.

The MitM Stealtooth attack easily achieves session hijacking including link key overwriting through the automatic pairing mode vulnerabilities we newly discovered. Furthermore, by combining with the existing Breaktooth attack, it abuses the vulnerabilities of Sleep mode that causes temporary Bluetooth disconnections without attacker intervention or victim operations, completely eliminating the jamming process from attack execution. The MitM Stealtooth attack provides a new attack vector against Bluetooth.

\section{Conclusion}

This paper unveils novel vulnerabilities of Bluetooth automatic pairing modes that enables completely silent device link key overwriting. We demonstrate how attackers can abuse the automatic pairing functions implemented in commercial Bluetooth devices to establish malicious connections without any user awareness or specialized equipment.

Our Stealtooth attack leverages the inherent behavior of commercial devices that automatically transition to pairing mode under specific conditions. We also extend the Stealtooth attack into a MitM attack, called MitM Stealtooth attacks, by combining the attack with existing power-saving mode techniques, enabling attackers to intercept, modify, and relay communications between victims.

We tested our attacks against 10 commercial Bluetooth devices from major manufacturers, including Sony, Anker, Google, and Xiaomi, and demonstrated the severe impact of the new vulnerabilities across various device chipsets, and vendors. Our practical implementation using only commodity hardware and open-source software highlights the real-world applicability of the attacks.

To address the critical impact of these vulnerabilities, we propose both device-level and protocol-level defenses. At the device level, we recommend enhanced user notification systems and stricter timeout mechanisms for automatic pairing. At the protocol level, we advocate for standardized automatic pairing guidelines and improved validation mechanisms for link key replacement scenarios.

Our findings reveal a critical tension between security and usability in wireless communication systems. While automatic pairing provides undeniable convenience benefits, current implementations inadequately consider security implications, creating systematic vulnerabilities rather than isolated implementation flaws.

\begin{acks}
This work is in part conducted under the ``Research and development on new generation cryptography for secure wireless communication services'' contract 
for the ``Research and Development for Expansion of Radio Wave Resources (JPJ000254)'', which is supported by the Ministry of Internal Affairs and Communications, Japan.
\end{acks}
\bibliographystyle{ACM-Reference-Format}
\bibliography{refs}


\begin{thebibliography}{33}


\ifx \showCODEN    \undefined \def \showCODEN     #1{\unskip}     \fi
\ifx \showISBNx    \undefined \def \showISBNx     #1{\unskip}     \fi
\ifx \showISBNxiii \undefined \def \showISBNxiii  #1{\unskip}     \fi
\ifx \showISSN     \undefined \def \showISSN      #1{\unskip}     \fi
\ifx \showLCCN     \undefined \def \showLCCN      #1{\unskip}     \fi
\ifx \shownote     \undefined \def \shownote      #1{#1}          \fi
\ifx \showarticletitle \undefined \def \showarticletitle #1{#1}   \fi
\ifx \showURL      \undefined \def \showURL       {\relax}        \fi
\providecommand\bibfield[2]{#2}
\providecommand\bibinfo[2]{#2}
\providecommand\natexlab[1]{#1}
\providecommand\showeprint[2][]{arXiv:#2}

\bibitem[Ai et~al\mbox{.}(2022)]%
        {ccs2022blacktooth}
\bibfield{author}{\bibinfo{person}{Mingrui Ai}, \bibinfo{person}{Kaiping Xue}, \bibinfo{person}{Bo Luo}, \bibinfo{person}{Lutong Chen}, \bibinfo{person}{Nenghai Yu}, \bibinfo{person}{Qibin Sun}, {and} \bibinfo{person}{Feng Wu}.} \bibinfo{year}{2022}\natexlab{}.
\newblock \showarticletitle{Blacktooth: Breaking through the Defense of Bluetooth in Silence}. In \bibinfo{booktitle}{\emph{Proceedings of the 2022 ACM SIGSAC Conference on Computer and Communications Security}} (Los Angeles, CA, USA) \emph{(\bibinfo{series}{CCS '22})}. \bibinfo{publisher}{Association for Computing Machinery}, \bibinfo{address}{New York, NY, USA}, \bibinfo{pages}{55–68}.
\newblock
\showISBNx{9781450394505}
\href{https://doi.org/10.1145/3548606.3560668}{doi:\nolinkurl{10.1145/3548606.3560668}}


\bibitem[Aminanto et~al\mbox{.}(2018)]%
        {ieeetifs2018wifi}
\bibfield{author}{\bibinfo{person}{Muhamad~Erza Aminanto}, \bibinfo{person}{Rakyong Choi}, \bibinfo{person}{Harry~Chandra Tanuwidjaja}, \bibinfo{person}{Paul~D. Yoo}, {and} \bibinfo{person}{Kwangjo Kim}.} \bibinfo{year}{2018}\natexlab{}.
\newblock \showarticletitle{Deep Abstraction and Weighted Feature Selection for Wi-Fi Impersonation Detection}.
\newblock \bibinfo{journal}{\emph{IEEE Transactions on Information Forensics and Security}} \bibinfo{volume}{13}, \bibinfo{number}{3} (\bibinfo{year}{2018}), \bibinfo{pages}{621--636}.
\newblock
\href{https://doi.org/10.1109/TIFS.2017.2762828}{doi:\nolinkurl{10.1109/TIFS.2017.2762828}}


\bibitem[Antonioli(2023)]%
        {ccs2023bluffs}
\bibfield{author}{\bibinfo{person}{Daniele Antonioli}.} \bibinfo{year}{2023}\natexlab{}.
\newblock \showarticletitle{BLUFFS: Bluetooth Forward and Future Secrecy Attacks and Defenses}. In \bibinfo{booktitle}{\emph{Proceedings of the 2023 ACM SIGSAC Conference on Computer and Communications Security}} (Copenhagen, Denmark) \emph{(\bibinfo{series}{CCS '23})}. \bibinfo{publisher}{Association for Computing Machinery}, \bibinfo{address}{New York, NY, USA}, \bibinfo{pages}{636–650}.
\newblock
\showISBNx{9798400700507}
\href{https://doi.org/10.1145/3576915.3623066}{doi:\nolinkurl{10.1145/3576915.3623066}}


\bibitem[Antonioli and Payer(2022)]%
        {spw2022vehicles}
\bibfield{author}{\bibinfo{person}{Daniele Antonioli} {and} \bibinfo{person}{Mathias Payer}.} \bibinfo{year}{2022}\natexlab{}.
\newblock \showarticletitle{On the Insecurity of Vehicles Against Protocol-Level Bluetooth Threats}. In \bibinfo{booktitle}{\emph{2022 IEEE Security and Privacy Workshops (SPW)}}. \bibinfo{pages}{353--362}.
\newblock
\href{https://doi.org/10.1109/SPW54247.2022.9833886}{doi:\nolinkurl{10.1109/SPW54247.2022.9833886}}


\bibitem[Antonioli et~al\mbox{.}(2020)]%
        {sp2020bias}
\bibfield{author}{\bibinfo{person}{Daniele Antonioli}, \bibinfo{person}{Nils~Ole Tippenhauer}, {and} \bibinfo{person}{Kasper Rasmussen}.} \bibinfo{year}{2020}\natexlab{}.
\newblock \showarticletitle{BIAS: Bluetooth Impersonation AttackS}. In \bibinfo{booktitle}{\emph{2020 IEEE Symposium on Security and Privacy (SP)}}. \bibinfo{pages}{549--562}.
\newblock
\href{https://doi.org/10.1109/SP40000.2020.00093}{doi:\nolinkurl{10.1109/SP40000.2020.00093}}


\bibitem[Antonioli et~al\mbox{.}(2019)]%
        {usenix2019knob}
\bibfield{author}{\bibinfo{person}{Daniele Antonioli}, \bibinfo{person}{Nils~Ole Tippenhauer}, {and} \bibinfo{person}{Kasper~B. Rasmussen}.} \bibinfo{year}{2019}\natexlab{}.
\newblock \showarticletitle{The {KNOB} is Broken: Exploiting Low Entropy in the Encryption Key Negotiation Of Bluetooth {BR/EDR}}. In \bibinfo{booktitle}{\emph{28th USENIX Security Symposium (USENIX Security 19)}}. \bibinfo{publisher}{USENIX Association}, \bibinfo{address}{Santa Clara, CA}, \bibinfo{pages}{1047--1061}.
\newblock
\showISBNx{978-1-939133-06-9}
\urldef\tempurl%
\url{https://www.usenix.org/conference/usenixsecurity19/presentation/antonioli}
\showURL{%
\tempurl}


\bibitem[Biham and Neumann(2019)]%
        {sac2019curve}
\bibfield{author}{\bibinfo{person}{Eli Biham} {and} \bibinfo{person}{Lior Neumann}.} \bibinfo{year}{2019}\natexlab{}.
\newblock \showarticletitle{Breaking the Bluetooth Pairing – The Fixed Coordinate Invalid Curve Attack}. In \bibinfo{booktitle}{\emph{Selected Areas in Cryptography – SAC 2019: 26th International Conference, Waterloo, ON, Canada, August 12–16, 2019, Revised Selected Papers}} (Waterloo, ON, Canada). \bibinfo{publisher}{Springer-Verlag}, \bibinfo{address}{Berlin, Heidelberg}, \bibinfo{pages}{250–273}.
\newblock
\showISBNx{978-3-030-38470-8}
\href{https://doi.org/10.1007/978-3-030-38471-5_11}{doi:\nolinkurl{10.1007/978-3-030-38471-5_11}}


\bibitem[Blancaflor et~al\mbox{.}(2025)]%
        {springer2025hcitool}
\bibfield{author}{\bibinfo{person}{Eric Blancaflor}, \bibinfo{person}{Harold~Kobe Billo}, \bibinfo{person}{John~Michael Dignadice}, \bibinfo{person}{Philip Domondon}, \bibinfo{person}{Mico~Ruiz Linco}, {and} \bibinfo{person}{Christie Valero}.} \bibinfo{year}{2025}\natexlab{}.
\newblock \showarticletitle{Bluetooth Simulated Reconnaissance Attack Through the Use of HCITool: A Case Study}. In \bibinfo{booktitle}{\emph{2nd International Conference on Cloud Computing and Computer Networks}}, \bibfield{editor}{\bibinfo{person}{Lei Meng}} (Ed.). \bibinfo{publisher}{Springer Nature Switzerland}, \bibinfo{address}{Cham}, \bibinfo{pages}{133--143}.
\newblock
\showISBNx{978-3-031-78131-5}


\bibitem[Che et~al\mbox{.}(2024)]%
        {ccs2024bluewswat}
\bibfield{author}{\bibinfo{person}{Xijia Che}, \bibinfo{person}{Yi He}, \bibinfo{person}{Xuewei Feng}, \bibinfo{person}{Kun Sun}, \bibinfo{person}{Ke Xu}, {and} \bibinfo{person}{Qi Li}.} \bibinfo{year}{2024}\natexlab{}.
\newblock \showarticletitle{BlueSWAT: A Lightweight State-Aware Security Framework for Bluetooth Low Energy}. In \bibinfo{booktitle}{\emph{Proceedings of the 2024 on ACM SIGSAC Conference on Computer and Communications Security}} (Salt Lake City, UT, USA) \emph{(\bibinfo{series}{CCS '24})}. \bibinfo{publisher}{Association for Computing Machinery}, \bibinfo{address}{New York, NY, USA}, \bibinfo{pages}{2087–2101}.
\newblock
\showISBNx{9798400706363}
\href{https://doi.org/10.1145/3658644.3670397}{doi:\nolinkurl{10.1145/3658644.3670397}}


\bibitem[Fischlin and Sanina(2024)]%
        {ccs2024fake}
\bibfield{author}{\bibinfo{person}{Marc Fischlin} {and} \bibinfo{person}{Olga Sanina}.} \bibinfo{year}{2024}\natexlab{}.
\newblock \showarticletitle{Fake It till You Make It: Enhancing Security of Bluetooth Secure Connections via Deferrable Authentication}. In \bibinfo{booktitle}{\emph{Proceedings of the 2024 on ACM SIGSAC Conference on Computer and Communications Security}} (Salt Lake City, UT, USA) \emph{(\bibinfo{series}{CCS '24})}. \bibinfo{publisher}{Association for Computing Machinery}, \bibinfo{address}{New York, NY, USA}, \bibinfo{pages}{4762–4776}.
\newblock
\showISBNx{9798400706363}
\href{https://doi.org/10.1145/3658644.3670360}{doi:\nolinkurl{10.1145/3658644.3670360}}


\bibitem[Garbelini et~al\mbox{.}(2022)]%
        {usenix2022braktooth}
\bibfield{author}{\bibinfo{person}{Matheus~E. Garbelini}, \bibinfo{person}{Vaibhav Bedi}, \bibinfo{person}{Sudipta Chattopadhyay}, \bibinfo{person}{Sumei Sun}, {and} \bibinfo{person}{Ernest Kurniawan}.} \bibinfo{year}{2022}\natexlab{}.
\newblock \showarticletitle{{BrakTooth}: Causing Havoc on Bluetooth Link Manager via Directed Fuzzing}. In \bibinfo{booktitle}{\emph{31st USENIX Security Symposium (USENIX Security 22)}}. \bibinfo{publisher}{USENIX Association}, \bibinfo{address}{Boston, MA}, \bibinfo{pages}{1025--1042}.
\newblock
\showISBNx{978-1-939133-31-1}
\urldef\tempurl%
\url{https://www.usenix.org/conference/usenixsecurity22/presentation/garbelini}
\showURL{%
\tempurl}


\bibitem[Gore et~al\mbox{.}(2019)]%
        {icondsc2019ble}
\bibfield{author}{\bibinfo{person}{Rahul~N. Gore}, \bibinfo{person}{Himashri Kour}, \bibinfo{person}{Mihit Gandhi}, \bibinfo{person}{Deepaknath Tandur}, {and} \bibinfo{person}{Anitha Varghese}.} \bibinfo{year}{2019}\natexlab{}.
\newblock \showarticletitle{Bluetooth based Sensor Monitoring in Industrial IoT Plants}. In \bibinfo{booktitle}{\emph{2019 International Conference on Data Science and Communication (IconDSC)}}. \bibinfo{pages}{1--6}.
\newblock
\href{https://doi.org/10.1109/IconDSC.2019.8816906}{doi:\nolinkurl{10.1109/IconDSC.2019.8816906}}


\bibitem[Halevi and Saxena(2013)]%
        {ieeetifs2013pairing}
\bibfield{author}{\bibinfo{person}{Tzipora Halevi} {and} \bibinfo{person}{Nitesh Saxena}.} \bibinfo{year}{2013}\natexlab{}.
\newblock \showarticletitle{Acoustic Eavesdropping Attacks on Constrained Wireless Device Pairing}.
\newblock \bibinfo{journal}{\emph{IEEE Transactions on Information Forensics and Security}} \bibinfo{volume}{8}, \bibinfo{number}{3} (\bibinfo{year}{2013}), \bibinfo{pages}{563--577}.
\newblock
\href{https://doi.org/10.1109/TIFS.2013.2247758}{doi:\nolinkurl{10.1109/TIFS.2013.2247758}}


\bibitem[Jangid et~al\mbox{.}(2023)]%
        {ndss2023pep}
\bibfield{author}{\bibinfo{person}{Mohit Jangid}, \bibinfo{person}{Yue Zhang}, {and} \bibinfo{person}{Zhiqiang Lin}.} \bibinfo{year}{2023}\natexlab{}.
\newblock \showarticletitle{Extrapolating Formal Analysis to Uncover Attacks in Bluetooth Passkey Entry Pairing}. In \bibinfo{booktitle}{\emph{2023, Network and Distributed System Security Symposium (NDSS)}}.
\newblock
\href{https://doi.org/10.14722/ndss.2023.23119}{doi:\nolinkurl{10.14722/ndss.2023.23119}}


\bibitem[Kimura et~al\mbox{.}(2024)]%
        {cryptoeprint2024breaktooth}
\bibfield{author}{\bibinfo{person}{Keiichiro Kimura}, \bibinfo{person}{Hiroki Kuzuno}, \bibinfo{person}{Yoshiaki Shiraishi}, {and} \bibinfo{person}{Masakatu Morii}.} \bibinfo{year}{2024}\natexlab{}.
\newblock \bibinfo{title}{Breaktooth: Breaking Security and Privacy in Bluetooth Power-Saving Mode}.
\newblock \bibinfo{howpublished}{Cryptology {ePrint} Archive, Paper 2024/900}.
\newblock
\urldef\tempurl%
\url{https://eprint.iacr.org/2024/900}
\showURL{%
\tempurl}


\bibitem[Koh et~al\mbox{.}(2022)]%
        {dsn2022blap}
\bibfield{author}{\bibinfo{person}{Changseok Koh}, \bibinfo{person}{Jonghoon Kwon}, {and} \bibinfo{person}{Junbeom Hur}.} \bibinfo{year}{2022}\natexlab{}.
\newblock \showarticletitle{BLAP: Bluetooth Link Key Extraction and Page Blocking Attacks}. In \bibinfo{booktitle}{\emph{2022 52nd Annual IEEE/IFIP International Conference on Dependable Systems and Networks (DSN)}}. \bibinfo{pages}{227--238}.
\newblock
\href{https://doi.org/10.1109/DSN53405.2022.00033}{doi:\nolinkurl{10.1109/DSN53405.2022.00033}}


\bibitem[Koulouras et~al\mbox{.}(2025)]%
        {sensor2025ble}
\bibfield{author}{\bibinfo{person}{Grigorios Koulouras}, \bibinfo{person}{Stylianos Katsoulis}, {and} \bibinfo{person}{Fotios Zantalis}.} \bibinfo{year}{2025}\natexlab{}.
\newblock \showarticletitle{Evolution of Bluetooth Technology: BLE in the IoT Ecosystem}.
\newblock \bibinfo{journal}{\emph{Sensors}} \bibinfo{volume}{25}, \bibinfo{number}{4} (\bibinfo{year}{2025}).
\newblock
\showISSN{1424-8220}
\href{https://doi.org/10.3390/s25040996}{doi:\nolinkurl{10.3390/s25040996}}


\bibitem[Mackensen et~al\mbox{.}(2012)]%
        {sensor2012ble}
\bibfield{author}{\bibinfo{person}{Elke Mackensen}, \bibinfo{person}{Matthias Lai}, {and} \bibinfo{person}{Thomas~M. Wendt}.} \bibinfo{year}{2012}\natexlab{}.
\newblock \showarticletitle{Bluetooth Low Energy (BLE) based wireless sensors}. In \bibinfo{booktitle}{\emph{SENSORS, 2012 IEEE}}. \bibinfo{pages}{1--4}.
\newblock
\href{https://doi.org/10.1109/ICSENS.2012.6411303}{doi:\nolinkurl{10.1109/ICSENS.2012.6411303}}


\bibitem[man page(2024)]%
        {l2ping}
\bibfield{author}{\bibinfo{person}{Linux man page}.} \bibinfo{year}{2002-2024}\natexlab{}.
\newblock \bibinfo{title}{l2ping(1)}.
\newblock \bibinfo{howpublished}{\url{https://linux.die.net/man/1/l2ping}}.
\newblock
\newblock
\shownote{Accessed: 2025-06-01}.


\bibitem[Pušnik et~al\mbox{.}(2020)]%
        {sensor2020ble}
\bibfield{author}{\bibinfo{person}{Maja Pušnik}, \bibinfo{person}{Mitja Galun}, {and} \bibinfo{person}{Boštjan Šumak}.} \bibinfo{year}{2020}\natexlab{}.
\newblock \showarticletitle{Improved Bluetooth Low Energy Sensor Detection for Indoor Localization Services}.
\newblock \bibinfo{journal}{\emph{Sensors}} \bibinfo{volume}{20}, \bibinfo{number}{8} (\bibinfo{year}{2020}).
\newblock
\showISSN{1424-8220}
\href{https://doi.org/10.3390/s20082336}{doi:\nolinkurl{10.3390/s20082336}}


\bibitem[Shelke et~al\mbox{.}(2024)]%
        {secrypt2024bluedos}
\bibfield{author}{\bibinfo{person}{Poonam Shelke}, \bibinfo{person}{Saurav Gupta}, {and} \bibinfo{person}{Sukumar Nandi}.} \bibinfo{year}{2024}\natexlab{}.
\newblock \showarticletitle{BlueDoS: A Novel Approach to Perform and Analyse DoS Attacks on Bluetooth Devices}. In \bibinfo{booktitle}{\emph{Proceedings of the 21st International Conference on Security and Cryptography - Volume 1: SECRYPT}}. INSTICC, \bibinfo{publisher}{SciTePress}, \bibinfo{pages}{838--843}.
\newblock
\showISBNx{978-989-758-709-2}
\href{https://doi.org/10.5220/0012845700003767}{doi:\nolinkurl{10.5220/0012845700003767}}


\bibitem[Shen et~al\mbox{.}(2022)]%
        {ieeetifs2022lora}
\bibfield{author}{\bibinfo{person}{Guanxiong Shen}, \bibinfo{person}{Junqing Zhang}, \bibinfo{person}{Alan Marshall}, {and} \bibinfo{person}{Joseph~R. Cavallaro}.} \bibinfo{year}{2022}\natexlab{}.
\newblock \showarticletitle{Towards Scalable and Channel-Robust Radio Frequency Fingerprint Identification for LoRa}.
\newblock \bibinfo{journal}{\emph{IEEE Transactions on Information Forensics and Security}}  \bibinfo{volume}{17} (\bibinfo{year}{2022}), \bibinfo{pages}{774--787}.
\newblock
\href{https://doi.org/10.1109/TIFS.2022.3152404}{doi:\nolinkurl{10.1109/TIFS.2022.3152404}}


\bibitem[SIG(2023)]%
        {bluetoothsig2023market}
\bibfield{author}{\bibinfo{person}{Bluetooth SIG}.} \bibinfo{year}{2023}\natexlab{}.
\newblock \bibinfo{title}{2023 Bluetooth® Market Update}.
\newblock \bibinfo{howpublished}{\url{https://www.bluetooth.com/2023-market-update/}}.
\newblock
\newblock
\shownote{Accessed: 2025-06-01}.


\bibitem[SIG(2024)]%
        {bluetoothsig2024market}
\bibfield{author}{\bibinfo{person}{Bluetooth SIG}.} \bibinfo{year}{2024}\natexlab{}.
\newblock \bibinfo{title}{2024 Bluetooth® Market Update}.
\newblock \bibinfo{howpublished}{\url{https://www.bluetooth.com/2024-market-update/}}.
\newblock
\newblock
\shownote{Accessed: 2025-06-01}.


\bibitem[SIG(2025)]%
        {bluetoothsig2025market}
\bibfield{author}{\bibinfo{person}{Bluetooth SIG}.} \bibinfo{year}{2025}\natexlab{}.
\newblock \bibinfo{title}{2025 Bluetooth® Market Update}.
\newblock \bibinfo{howpublished}{\url{https://www.bluetooth.com/2025-market-update/}}.
\newblock
\newblock
\shownote{Accessed: 2025-06-01}.


\bibitem[Sun et~al\mbox{.}(2018)]%
        {puc2018ssp}
\bibfield{author}{\bibinfo{person}{Da-Zhi Sun}, \bibinfo{person}{Yi Mu}, {and} \bibinfo{person}{Willy Susilo}.} \bibinfo{year}{2018}\natexlab{}.
\newblock \showarticletitle{Man-in-the-middle attacks on Secure Simple Pairing in Bluetooth standard V5.0 and its countermeasure}.
\newblock \bibinfo{journal}{\emph{Personal Ubiquitous Comput.}} \bibinfo{volume}{22}, \bibinfo{number}{1} (\bibinfo{date}{Feb.} \bibinfo{year}{2018}), \bibinfo{pages}{55–67}.
\newblock
\showISSN{1617-4909}
\href{https://doi.org/10.1007/s00779-017-1081-6}{doi:\nolinkurl{10.1007/s00779-017-1081-6}}


\bibitem[Sun and Sun(2019)]%
        {mdpi2019ssp}
\bibfield{author}{\bibinfo{person}{Da-Zhi Sun} {and} \bibinfo{person}{Li Sun}.} \bibinfo{year}{2019}\natexlab{}.
\newblock \showarticletitle{On Secure Simple Pairing in Bluetooth Standard v5.0-Part I: Authenticated Link Key Security and Its Home Automation and Entertainment Applications}.
\newblock \bibinfo{journal}{\emph{Sensors}} \bibinfo{volume}{19}, \bibinfo{number}{5} (\bibinfo{year}{2019}).
\newblock
\showISSN{1424-8220}
\href{https://doi.org/10.3390/s19051158}{doi:\nolinkurl{10.3390/s19051158}}


\bibitem[Tschirschnitz et~al\mbox{.}(2025)]%
        {ndss2025mc}
\bibfield{author}{\bibinfo{person}{Maximilian Tschirschnitz}, \bibinfo{person}{Ludwig Peuckert}, \bibinfo{person}{Moritz Buhl}, {and} \bibinfo{person}{Jens Grossklags}.} \bibinfo{year}{2025}\natexlab{}.
\newblock \showarticletitle{Rediscovering Method Confusion in Proposed Security Fixes for Bluetooth}. In \bibinfo{booktitle}{\emph{2025, Network and Distributed System Security Symposium (NDSS)}}.
\newblock
\href{https://doi.org/10.14722/ndss.2025.240310}{doi:\nolinkurl{10.14722/ndss.2025.240310}}


\bibitem[Tucker et~al\mbox{.}(2023)]%
        {sp2023bluesclues}
\bibfield{author}{\bibinfo{person}{Tyler Tucker}, \bibinfo{person}{Hunter Searle}, \bibinfo{person}{Kevin Butler}, {and} \bibinfo{person}{Patrick Traynor}.} \bibinfo{year}{2023}\natexlab{}.
\newblock \showarticletitle{Blue’s Clues: Practical Discovery of Non-Discoverable Bluetooth Devices}. In \bibinfo{booktitle}{\emph{2023 IEEE Symposium on Security and Privacy (SP)}}. \bibinfo{pages}{3098--3112}.
\newblock
\href{https://doi.org/10.1109/SP46215.2023.10179358}{doi:\nolinkurl{10.1109/SP46215.2023.10179358}}


\bibitem[von Tschirschnitz et~al\mbox{.}(2021)]%
        {sp2021mc}
\bibfield{author}{\bibinfo{person}{Maximilian von Tschirschnitz}, \bibinfo{person}{Ludwig Peuckert}, \bibinfo{person}{Fabian Franzen}, {and} \bibinfo{person}{Jens Grossklags}.} \bibinfo{year}{2021}\natexlab{}.
\newblock \showarticletitle{Method Confusion Attack on Bluetooth Pairing}. In \bibinfo{booktitle}{\emph{2021 IEEE Symposium on Security and Privacy (SP)}}. \bibinfo{pages}{1332--1347}.
\newblock
\href{https://doi.org/10.1109/SP40001.2021.00013}{doi:\nolinkurl{10.1109/SP40001.2021.00013}}


\bibitem[Wu et~al\mbox{.}(2020)]%
        {woot2020blesa}
\bibfield{author}{\bibinfo{person}{Jianliang Wu}, \bibinfo{person}{Yuhong Nan}, \bibinfo{person}{Vireshwar Kumar}, \bibinfo{person}{Dave~(Jing) Tian}, \bibinfo{person}{Antonio Bianchi}, \bibinfo{person}{Mathias Payer}, {and} \bibinfo{person}{Dongyan Xu}.} \bibinfo{year}{2020}\natexlab{}.
\newblock \showarticletitle{{BLESA}: Spoofing Attacks against Reconnections in Bluetooth Low Energy}. In \bibinfo{booktitle}{\emph{14th USENIX Workshop on Offensive Technologies (WOOT 20)}}. \bibinfo{publisher}{USENIX Association}.
\newblock
\urldef\tempurl%
\url{https://www.usenix.org/conference/woot20/presentation/wu}
\showURL{%
\tempurl}


\bibitem[Wu et~al\mbox{.}(2022)]%
        {sp2022btdesign}
\bibfield{author}{\bibinfo{person}{Jianliang Wu}, \bibinfo{person}{Ruoyu Wu}, \bibinfo{person}{Dongyan Xu}, \bibinfo{person}{Dave~Jing Tian}, {and} \bibinfo{person}{Antonio Bianchi}.} \bibinfo{year}{2022}\natexlab{}.
\newblock \showarticletitle{Formal Model-Driven Discovery of Bluetooth Protocol Design Vulnerabilities}. In \bibinfo{booktitle}{\emph{2022 IEEE Symposium on Security and Privacy (SP)}}. \bibinfo{pages}{2285--2303}.
\newblock
\href{https://doi.org/10.1109/SP46214.2022.9833777}{doi:\nolinkurl{10.1109/SP46214.2022.9833777}}


\bibitem[Yüksel et~al\mbox{.}(2022)]%
        {ssrn2022l2ping}
\bibfield{author}{\bibinfo{person}{Tuğrul Yüksel}, \bibinfo{person}{Ömer Aydın}, {and} \bibinfo{person}{Gökhan Dalkılıç}.} \bibinfo{year}{2022}\natexlab{}.
\newblock \showarticletitle{Performing DoS Attacks on Bluetooth Devices Paired with Google Home Mini}.
\newblock \bibinfo{journal}{\emph{SSRN Electronic Journal}}  \bibinfo{volume}{18} (\bibinfo{date}{01} \bibinfo{year}{2022}), \bibinfo{pages}{53--58}.
\newblock
\href{https://doi.org/10.2139/ssrn.4171322}{doi:\nolinkurl{10.2139/ssrn.4171322}}


\end{thebibliography}

\end{document}